\documentstyle[12pt,epsfig]{article}
\newcommand{\be}{\begin{equation}}
\newcommand{\ee}{\end{equation}}
\newcommand{\bea}{\begin{eqnarray}}
\newcommand{\eea}{\end{eqnarray}}
\newcommand{\eqn}[1]{(\ref{#1})}
\newcommand{\pp}{~~~.}
\newcommand{\vv}{~~~,}
\newcommand{\ds}{\displaystyle}
\newcommand{\nn}{\nonumber}

\newcommand{\rt}{\rightarrow}

\newcommand{\pe}{{\phi_e}}
\newcommand{\p}{{\rm p}}

\newcommand{\hnb}{\hat{n}_B}
\newcommand{\hH}{\widehat{H}}
\newcommand{\hmu}{\widehat{M}_u}
\newcommand{\hdmi}{\Delta \widehat{M}_i}
\newcommand{\hdmj}{\Delta \widehat{M}_j}
\newcommand{\hgi}{\widehat{\Gamma}_i}

\newcommand{\hrho}{\hat{\rho}}
\newcommand{\hp}{\hat{\rm p}}

\newcommand{\Nature}{{\it Nature\,}}

\newcommand{\ApJ}{{\it Astrophys. J.\,}}
\newcommand{\ApJS}{{\it Astrophys. J. Suppl.\,}}

\newcommand{\NP}{{\it Nucl. Phys.\,}}
\newcommand{\PR}{{\it Phys. Rev.\,}}
\newcommand{\PRL}{{\it Phys. Rev. Lett.\,}}

\newcommand{\PL}{{\it Phys. Lett.\,}}

\begin{document}
\setlength{\unitlength}{1mm}

\setlength{\unitlength}{1mm}
{\hfill $\begin{array}{r}
                \mbox{DSF 13/2000} \\
                \mbox{astro-ph/0005571}
\end{array}$}\vspace*{1cm}

\begin{center}
{\Large \bf The standard and degenerate primordial nucleosynthesis versus
recent experimental data}
\end{center}

\bigskip\bigskip

\begin{center}
{\bf S. Esposito}, {\bf G. Mangano}, {\bf G. Miele}, and {\bf O. Pisanti},
\end{center}

\vspace{.5cm}

\noindent
{\it Dipartimento di Scienze Fisiche, Universit\'{a} di Napoli "Federico
II", and INFN, Sezione di Napoli, Complesso Universitario di Monte
Sant'Angelo, Via Cintia, I-80126 Napoli, Italy}
\bigskip\bigskip\bigskip

\begin{abstract}
We report the results on Big Bang Nucleosynthesis (BBN) based on an updated
code, with accuracy of the order of $0.1 \%$ on $^4He$ abundance, compared
with the predictions of other recent similar analysis. We discuss the
compatibility of the theoretical results, for vanishing neutrino chemical
potentials, with the observational data. Bounds on the number of
relativistic neutrinos and baryon abundance are obtained by a likelihood
analysis. We also analyze the effect of large neutrino chemical potentials
on primordial nucleosynthesis, motivated by the recent results on the
Cosmic Microwave Background Radiation spectrum. The BBN exclusion plots for
electron neutrino chemical potential and the effective number of
relativistic neutrinos are reported. We find that the standard BBN seems to
be only marginally in agreement with the recent BOOMERANG and MAXIMA-1
results, while the agreement is much better for degenerate BBN scenarios
for large effective number of neutrinos, $N_\nu \sim 10$.
\end{abstract}
\vspace*{2cm}

\begin{center}
{\it PACS number(s): 98.80.Cq; 98.80.Ft}
\end{center}

\newpage
\baselineskip=.8cm

\section{Introduction}
\setcounter{equation}0

\noindent
The synthesis of light nuclei in the early universe, the Big Bang
Nucleosynthesis (BBN), represents one of the most striking evidence in
favour of standard cosmology, and since its proposal \cite{Wagoner}, it has
been extensively used as one of the best laboratories where to test
cosmological models and/or elementary particle physics. The appealing
feature of BBN is that, in its standard version, it relies on quite solid
theoretical grounds, which makes the predictions for $D$, $^3He$, $^4He$
and $^7Li$ abundances, whose primordial values are only partially modified
by the subsequent stellar activity, quite robust.

Recently, the experimental accuracy in measurements of the light
primordial nuclide abundances, mainly $^4He$, has been highly improved,
reaching a precision of the order of $1 \%$. Similar improvements have
been also obtained for both Deuterium ($D$) and $^7Li$ relative
abundances, $Y_D \equiv D/H$ and $Y_{^7Li} \equiv ~^7Li/H$, but,
unfortunately, the refinement of the experimental techniques does not yet
correspond to a clear picture of the primordial nuclide densities. This is
mainly due to an uncomplete understanding of systematic errors. In
particular, measuring the primordial $^4He$ mass fraction, $Y_p$, from
regression to zero metallicity in Blue Compact Galaxies, two independent
surveys obtained two results, a {\it low } value \cite{Steigman},
\be
Y_p^{(l)} = 0.234 {\pm} 0.003 \vv
\label{e:lowHe}
\ee
and a sensibly larger one \cite{Izotov},
\be
Y_p^{(h)} = 0.244 {\pm} 0.002 \vv
\label{e:highHe}
\ee
which are compatible at $2\sigma$ level only.

As in a recent analysis \cite{olive99}, we here adopt a more conservative
value, with a larger error (hereafter we always use 1$\sigma$ errors)
\be
Y_p = 0.238 {\pm} 0.005 \pp
\label{e:olive99}
\ee
A similar dichotomy holds in $D$ measurements as well, where the study of
distant Quasars Absorption line Systems (QAS), is thought to represent a
reliable way to estimate the primordial Deuterium. In this case,
observations in different QAS leads to the incompatible results
\cite{Songaila,Tytler,olive99}
\bea
Y_D^{(l)} &=& \left(3.4 {\pm} 0.3 \right) 10^{-5} \vv
\label{e:lowD} \\
Y_D^{(h)} &=& \left(2.0 {\pm} 0.5 \right) 10^{-4} \pp
\label{e:highD}
\eea
Finally, a reliable estimate for $^7Li$ primordial abundance is provided by
the {\it Spite plateau}, observed in the halo of POP II stars
\cite{Thorburn, Molaro}. The observations give the primordial abundance
\cite{Bonifacio},
\be
Y_{^7Li} =\left(1.73 {\pm} 0.21 \right) 10^{-10} \pp
\label{e:Li7}
\ee
From the theoretical point of view, the BBN predictions are obtained by
numerically solving a set of coupled Boltzmann equations, which trace the
abundances of the different nuclides in the framework of standard Big Bang
cosmology \cite{Kawano}. The collisional integrals of the above equations
contain all $n \leftrightarrow p $ weak reaction rates and a large nuclear
reaction network \cite{network}.

The increasing precision in measuring the primordial abundance has
recently pushed the theoretical community to make an effort to develop new
generation BBN codes \cite{Lopez,EMMP2}, with a comparable level of
accuracy. In a recent series of papers, in the framework of the standard
cosmological model, the present authors \cite{EMMP1} and other groups
\cite{Lopez} have performed a comprehensive and accurate analysis of all
the physical effects which influence $^4He$ mass fraction up to $0.1 \%$.

Since almost all neutrons present at the onset of nucleosynthesis are fixed
into $^4He$, its abundance is mainly function of the neutron versus proton
abundances, at the time of $n \leftrightarrow p$ weak interactions freeze
out, which takes place for $T \sim 1\, MeV$. To improve the accuracy on
the $^4He$ prediction it is demanding to reach an accuracy level of the
order of $1 \%$ in the estimate of the $n \leftrightarrow p$ rates, well
beyond the simple Born approximation. This has been performed by
considering a number of additional contributions, which, ordered according
to their relative weight, are the following:\\
i) electromagnetic radiative and Coulomb corrections
\cite{Sirlin,MarcianoSirlin};\\
ii) finite nucleon mass corrections \cite{Seckel, EMMP1};\\
iii) thermal radiative effects induced by the presence of a surrounding
plasma of $e^{\pm}$ and $\gamma$ \cite{FTQFT,EMMP1};\\
iv) corrections to the equation of state of the $e^{\pm}, \, \gamma$
plasma due to thermal mass renormalization \cite{stateeq,Lopez,EMMP2};\\
v) the residual neutrino coupling to the plasma during $e^+ e^-$
annihilation, affecting the neutrino to photon temperature ratio
\cite{residual,Lopez,EMMP2}.

Unfortunately, a similar systematic analysis of all corrections up to the
desired level of accuracy cannot be performed for the other input
parameters of the theory. The theoretical estimates do depend in fact on
the values of the neutron lifetime, as well as on several nuclear reaction
cross sections, which are poorly known in the energy range relevant for
BBN. This introduces a certain level of uncertainty on the light element
abundances. This aspect has been studied in two different approaches,
either using Monte Carlo methods \cite{Montecarlo} to sample the error
distributions of the relevant reaction cross sections, or, alternatively,
using a linear error propagation \cite{NuclearSigma}, with comparable
results.

In this paper, we further refine our previous predictions by using a new
version of our numerical code, where the full dependence of the BBN
equations on the electron chemical potential is accurately implemented.
Furthermore, for the {\it standard scenario} (vanishing neutrino chemical
potentials) we perform an accurate likelihood analysis in the space of the
two free parameters of the model, the effective number of neutrinos,
$N_\nu$, and the baryon to photon ratio $\eta$.

While the value of electron chemical potential is bounded, by neutrality,
by the value of $\eta$, this is not the case for neutrino--antineutrino
asymmetries which, in principle, can be quite large. The influence of the
neutrino chemical potentials on BBN predictions ({\it degenerate} BBN) has
been considered in the past \cite{KangSteigman}. One relevant aspect of
this analysis, which is worth stressing, is that degenerate BBN allows for
a better agreement with observations at values of $\eta$ larger than, say,
$6 {\cdot}10^{-10}$, while standard BBN prefers smaller values, $\eta \sim
(2 \div 6) {\cdot} 10^{-10}$. We will discuss this in detail in the paper.
This feature is particularly relevant in view of the recent analysis of the
BOOMERANG and MAXIMA-1 results \cite{Boomerang} on the acoustic peak of the
Cosmic Microwave Background Radiation (CMBR), which seems to favour a value
for the baryonic asymmetry $\eta$ of the order of $10^{-9}$, larger, as we
said, than what expected in the framework of standard BBN. This discrepancy
may be looked as a signal in favour of a large neutrino--antineutrino
asymmetry. It seems therefore demanding to re--analyze the degenerate
scenario, making profit of the now available more precise BBN codes. We
have performed this study, and the main result is that degenerate BBN and
CMBR data seems to be compatible for large values of the effective number
of neutrinos, $N_\nu \geq 10$ and $\eta \leq 10^{-9}$.

The paper is organized as follows. In section 2 we briefly describe the BBN
set of differential equations, recast in a suitable form for a numerical
solution. The light element abundances obtained from our code, for standard
BBN, are then presented in section 3 as a functions of $\tau_n$, $N_\nu$
and $\eta$. Section 4 is devoted to degenerate BBN and to the bounds on
neutrino chemical potentials coming from nucleosynthesis and CMBR data.
Finally, in section 4, we give our conclusions.

\section{The BBN set of equations}
\setcounter{equation}0

Consider $N_{nuc}$ species of nuclides, whose number densities, $n_i$, are
normalized with respect to the total number density of baryons,
$n_B$,\footnote{We will use the same notations of our previous paper
\cite{EMMP2}, which we refer to for further details.}
\be
X_i=\frac{n_i}{n_B} \quad\quad\quad i=1,..,N_{nuc} \pp
\ee
Alternatively, we will also make use in the following of the notation:
\bea
X_1 = X_n \vv &~~~~~ X_2 = X_p =X_H \vv &~~~~~ X_3 = X_D \vv ~~~~~
\nonumber \\
X_5 = X_{^3He} \vv &~~~~~ X_6 = X_{^4He} \vv &~~~~~ X_8 = X_{^7Li} \pp
\eea
The set of differential equations ruling primordial nucleosynthesis is
given by \cite{Wagoner,EMMP2}:
\bea
\frac{\dot{R}}{R}  &=& H = \sqrt{\frac{8\, \pi}{3\, M_P^2}~ \rho_T} \vv
\label{e:drdt} \\
\frac{\dot{n}_B}{n_B} &=& -\, 3\, H \vv
\label{e:dnbdt} \\
\dot{\rho_T} &=& -\, 3 \, H~ (\rho_T + \p_T) \vv
\label{e:drhodt} \\
\dot{X}_i &=& \sum_{j,k,l}\, N_i \left(  \Gamma_{kl \rt ij}\,
\frac{X_l^{N_l}\, X_k^{N_k}}{N_l!\, N_k !}  \; - \; \Gamma_{ij \rt kl}\,
\frac{X_i^{N_i}\, X_j^{N_j}}{N_i !\, N_j  !} \right) \equiv \Gamma_i(X_j)
\vv
\label{e:dXdt} \\
L (\frac{m_e}{T}, \pe) &=& \frac{n_B}{T^3}~ \sum_j Z_j\, X_j \vv
\label{e:charneut}
\eea
where $\rho_T$ and $\p_T$ denote the total energy density and pressure,
respectively,
\bea
\rho_T &=& \rho_\gamma + \rho_e + \rho_\nu + \rho_B \equiv \rho_{NB} +
\rho_B \vv \\
\p_T &=& \p_\gamma + \p_e + \p_\nu + \p_B \vv
\eea
$i,j,k,l=(1,..,N_{nuc})$, $Z_i$ is the charge number of the $i-$th nuclide,
and the function $L(z,y)$ is defined as
\be
L(z,y) \equiv \frac{1}{\pi^2} \int_z^\infty dx~x\, \sqrt{x^2-z^2}~ \left(
\frac{1}{e^{x-y}+1} - \frac{1}{e^{x+y}+1} \right) \pp
\ee
Eq.(\ref{e:drdt}) is the definition of the Hubble parameter, $H$, whereas
Eq.s (\ref{e:dnbdt}) and (\ref{e:drhodt}) state the total baryon number and
entropy conservation in the comoving volume, respectively. The set of
$N_{nuc}$ Boltzmann equations (\ref{e:dXdt}) describe the density evolution
of each nuclide specie, and finally Eq.(\ref{e:charneut}) states the
universe charge neutrality in terms of the electron chemical potential,
$\phi_e \equiv \mu_e/T$, with $T$ the temperature of $e^{\pm},\gamma$
plasma. Note that the neutrino energy density and pressure are included in
Eq.~\eqn{e:drhodt} only for $T \geq T_D$\footnote{We assume that all
neutrinos decouple at the same temperature, $T_D = 2.3\, MeV$
\cite{Enqvist}.}.

In a previous analysis, \cite{EMMP2}, we neglected the contribution of
$\pe$ in the BBN equations, since its effect results to be very small. In
this way we obtained a substantial simplification of the set of equations
(\ref{e:drdt})-(\ref{e:charneut}), since the unknown functions can be
reduced in this case to the $N_{nuc}+1$  ($\hat{h} \equiv n_B/T^3,~ X_j$).
We do here release this assumption and report the results of an improved
code where we take into complete account the electron chemical potential
evolution. We will see, however, that all changes on the final abundances
are of minor impact. In order to obtain the new equations, we note that it
is more convenient to follow the evolution of the $N_{nuc}+1$ unknown
functions $(\pe,~ X_j)$ in terms of the dimensionless variable $z=m_e/T$,
and to use Eq.~\eqn{e:charneut} to get $n_B$ as a function of $\pe$. The
new set of differential equations may be cast in the form
\be
\frac{d \pe}{dz}\, =\, \frac{1}{z}~ \frac{L~ E~ F + (z\, L_z - 3\, L)~
G}{L~ E~ \frac{\ds \delta \hrho_e}{\ds \delta \pe} - L_\pe~ G} \vv
\label{e:basic1}
\ee
\be
\frac{dX_i}{dz}\, =\, - \frac{\hgi}{z}~ \frac{L_\pe~ F + (z\, L_z - 3\, L)~
\frac{\ds \delta \hrho_e}{\ds \delta \pe}}{L~ E~ \frac{\ds \delta
\hrho_e}{\ds \delta \pe} - L_\pe~ G} \vv
\label{e:basic2}
\ee
where the functions $E$, $F$ and $G$ are given by
\bea
E(z,\, \pe,\, X_j) &=& 3\, \hH - \frac{ \ds \sum_i~ Z_i\, \hgi}{\ds \sum_j~
Z_j\, X_j} \vv
\label{e:funce} \\
F(z,\, \pe,\, X_j) &=& 4\, \hrho_{NB} + \frac32~ \hp_B - z\, \frac{\delta
\hrho_e}{\delta z} \vv
\label{e:funcg} \\
G(z,\, \pe,\, X_j) &=& 3\, \hH~ (\hrho_{NB} + \hp) + \frac{z\, L}{\sum_j
Z_j\, X_j}~ \sum_i \left( \hdmi + \frac{3}{2\, z} \right)~ \hgi \vv
\label{e:funcf}
\eea
and $H \equiv m_e\, \hH$, $n_B \equiv m_e^3~ \hnb$, $\Gamma_i \equiv m_e\,
\hgi$, $\rho_T \equiv T^4\,\hrho_T$, $\p_T \equiv T^4\, \hp_T$. Note that
by using Eq.~\eqn{e:charneut} and the previous definitions, it is possible
to express $\hrho_B$ and $\hp_B$ as functions of $z$, $\pe$, and $X_i$ only
\bea
\hrho_B &=& \frac{z\, L (z, \pe)}{\sum_j Z_j\, X_j}~ \left[ \hmu + \sum_j
\left( \hdmj\, +\, \frac{3}{2\, z} \right)\, X_j \right] \vv \\
\hp_B &=& \frac{L (z, \pe)}{\sum_j Z_j\, X_j}~ \sum_j X_j \pp
\eea
With $\hdmi$ and $\hmu$ we denote the i-th nuclide mass excess and the
atomic mass unit, respectively, normalized to $m_e$. In Appendix A we
report the partial derivative of $L$ with respect to $z$ and $\pe$, denoted
with $L_z$ and $L_{\pe}$, and the quantities $\hrho_e$, $\delta
\hrho_e/\delta z$ and $\delta \hrho_e/\delta \pe$ in a form which is
suitable for a BBN code implementation.

Eq.s~\eqn{e:basic1}-\eqn{e:basic2} are solved by imposing the following
initial conditions at $z_{in}= m_e/(10\, MeV)$:
\bea
\pe (z_{in}) &=& \pe^0 \vv \\
X_n (z_{in}) &=& \left(\exp\{\hat{q}\, z_{in}\}+1\right)^{-1} \vv
\quad\quad X_p (z_{in}) = \left(\exp\{-\hat{q}\, z_{in}\}+1\right)^{-1} \vv
\\
X_i (z_{in}) &=& \frac{g_i}{2}~ \left( \zeta(3) \sqrt{\frac{8}{\pi}}
\right)^{A_i-1} ~ A_i^\frac{3}{2}\, \left( \frac{m_e}{M_N z_{in}}
\right)^{\frac{3}{2} (A_i-1)} \eta^{A_i-1}\, X_p^{Z_i}\, X_n^{A_i-Z_i}  \nn
\\
&{\times}& \exp \left\{ \hat{B}_i \, z_{in} \right\}
\quad\quad\quad\quad\quad\quad\quad\quad\quad\quad
\mbox{with}~i=3,..,N_{nuc} \pp
\eea
In the previous equations $\hat{q}=(M_n- M_p)/m_e$, and the quantities
$A_i$ and $\hat{B}_i$ denote the atomic number and the binding energy of
the $i-$th nuclide normalized to electron mass, respectively. Finally
$\eta$ is, as usual, the baryon to photon number density ratio, and $\pe^0$
the solution of the implicit equation
\be
L (z_{in},\, \pe^0) = \frac{11}{4}~ \frac{2\, \zeta(3)}{\pi^2}~ \eta~
\sum_i Z_i\, X_i (z_{in}) \pp
\ee
The method of resolution of the BBN equations \eqn{e:basic1}-\eqn{e:basic2}
is the same applied in \cite{EMMP2}. It is the Backward Differentiation
Formulas with Newton's method, implemented in a NAG routine with adaptive
step-size (see \cite{EMMP2} for more details). We used the reduced network
of nuclear reactions, made of 25 reactions involving 9 nuclides, since the
use of the complete network affects the abundances for no more than $0.01
~\%$.

\section{Primordial abundances for standard BBN}
\setcounter{equation}0

The new BBN code has been used to produce the primordial abundances, in the
standard scenario, for different values of the input parameters, namely the
neutron lifetime $\tau_n$, the effective number of neutrinos $N_\nu$,
defined as
\be
\rho_\nu =  N_\nu \, \frac{7}{4} \, \frac{\pi^2}{30} \, T_\nu^4 \vv
\label{neff}
\ee
with $\rho_\nu$ the total neutrino energy densities, and the final baryon
to photon number density ratio $\eta$.

We consider the abundances relative to hydrogen for $D$, $^3He$, and
$^7Li$,
\be
Y_D = \frac{X_D}{X_H} \vv ~~~~~ Y_{^3He} = \frac{X_{^3He}}{X_H} \vv ~~~~~
Y_{^7Li} = \frac{X_{^7Li}}{X_H} \pp
\label{e:abund237}
\ee
In the case of $^4He$, the quantity usually defined as {\it mass fraction},
\be
Y_p = \frac{A_{^4He}~ X_{^4He}}{\sum_j A_j~ X_j} \vv
\label{e:abund4}
\ee
is rather the baryon number fraction, since $A_{^4He}=4$. Note that
expression \eqn{e:abund4} does not correspond to the true mass fraction,
obviously defined as
\be
Y_p^m = \frac{M_{^4He}~ X_{^4He}}{\sum_j M_j~ X_j} \pp
\ee
The difference between $Y_p$ and $Y_p^m$ are of the order of $1\%$ and thus
relevant for an accurate analysis like the one presented here. Since it is
customary to express the experimental value in terms of $Y_p$, see \eqn
{e:olive99}, we will consider this quantity for a comparison with
experimental data.

Table \ref{t:abund} shows the results obtained for $N_\nu = 3$, $\tau_n =
886.7\, s$, and $\eta = 5{\cdot} 10^{-10}$, compared with our previous
results \cite{EMMP2}. As one can see, the inclusion of the complete
evolution of $\pe$ modifies the final $^4He$ abundances for less than
$0.1\%$ and for few percent the other nuclides.

\begin{figure}
\epsfig{file=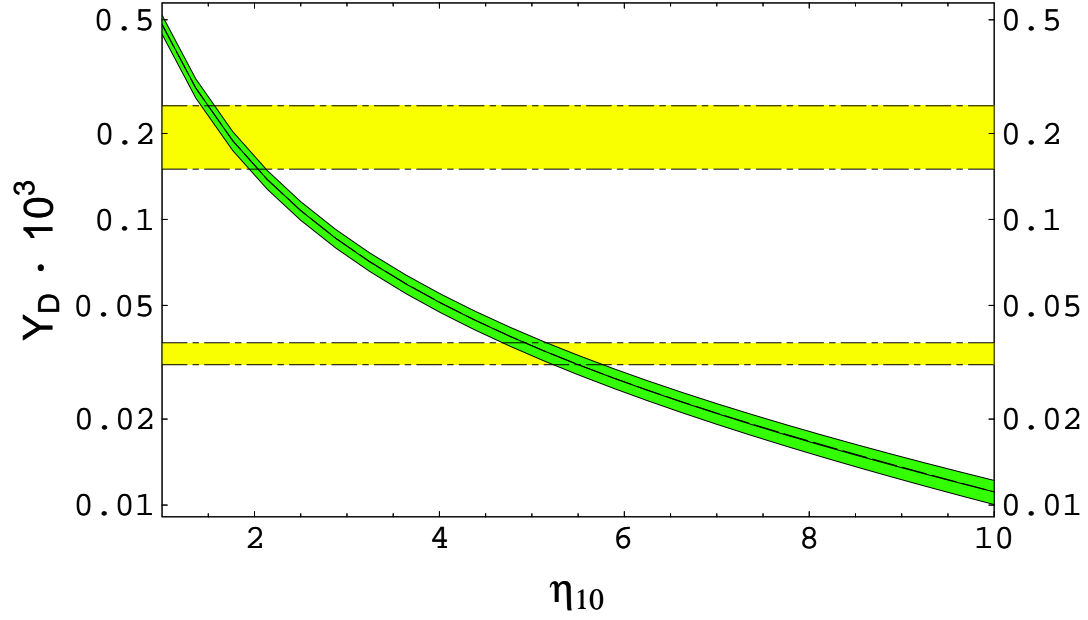,height=9cm}
\caption{The Deuterium abundance $Y_D$ \protect\eqn{e:abund237}, for
$N_\nu=3$, versus $\eta_{10} \equiv 10^{10} \eta$. The horizontal bands
correspond to the experimental determinations of Ref.s
\protect\cite{Songaila,Tytler}. The solid lines bound the theoretical
predictions at $1\sigma_{th}$. The dashed line, on this scale
indistinguishable from the central solid line, is the result from
\protect\cite{NuclearSigma}. No result is given for $D$ in
\protect\cite{Lopez}. The same notation has been applied in the following
Figures.}
\label{f:D}
\end{figure}

\begin{figure}
\epsfig{file=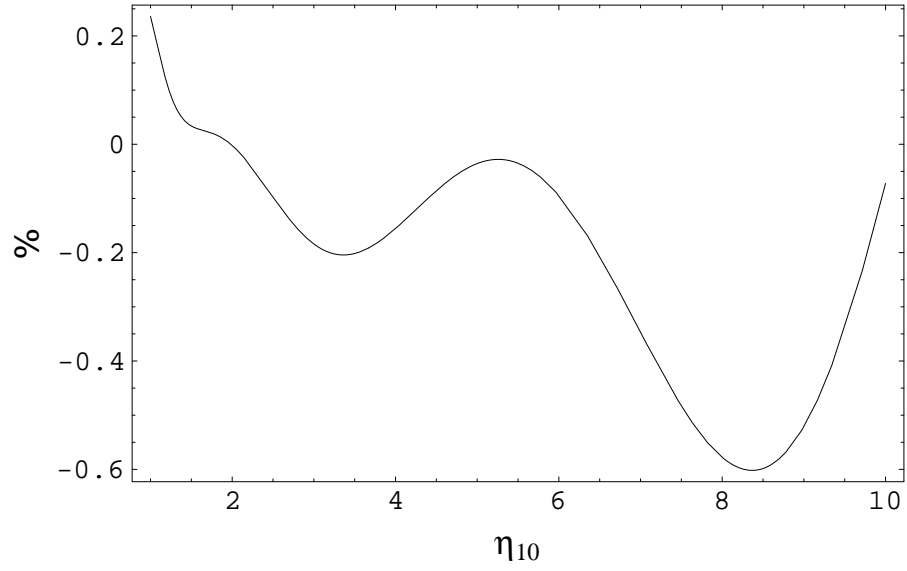,height=8cm}
\caption{Relative difference (in percent) for Deuterium abundance between
the results of the present work and the ones from
\protect\cite{NuclearSigma} in the case $N_\nu=3$.}
\label{f:compd}
\end{figure}

\begin{figure}
\epsfig{file=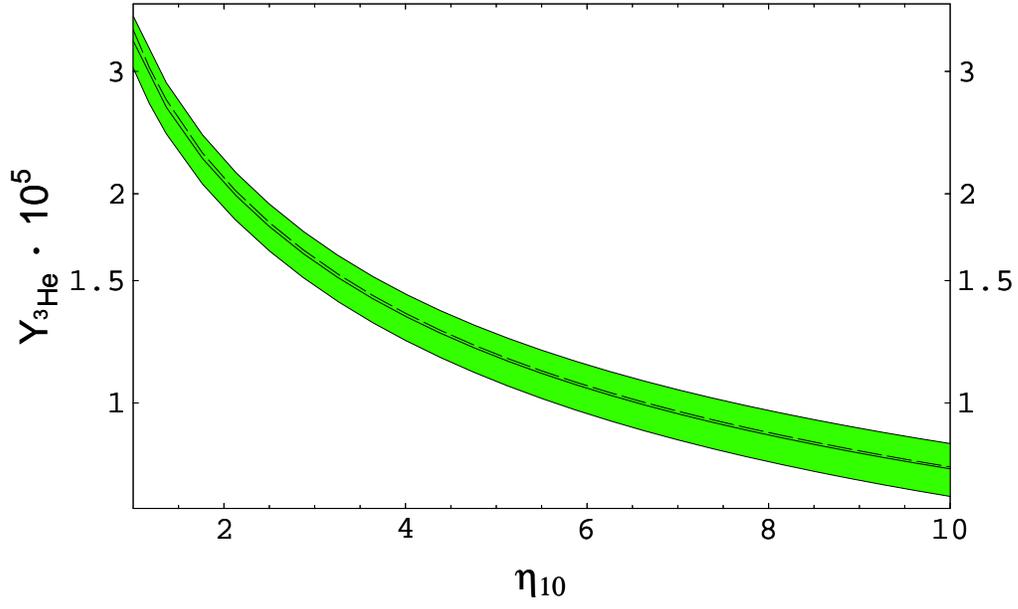,height=9cm}
\caption{The helium-3 abundance $Y_{^3He}$ \protect\eqn{e:abund237}, for
$N_\nu=3$, versus $\eta_{10}$. The dashed line is the result from
\protect\cite{NuclearSigma}. No result is given for $^3He$ in
\protect\cite{Lopez}.}
\label{f:He3}
\end{figure}

\begin{figure}
\epsfig{file=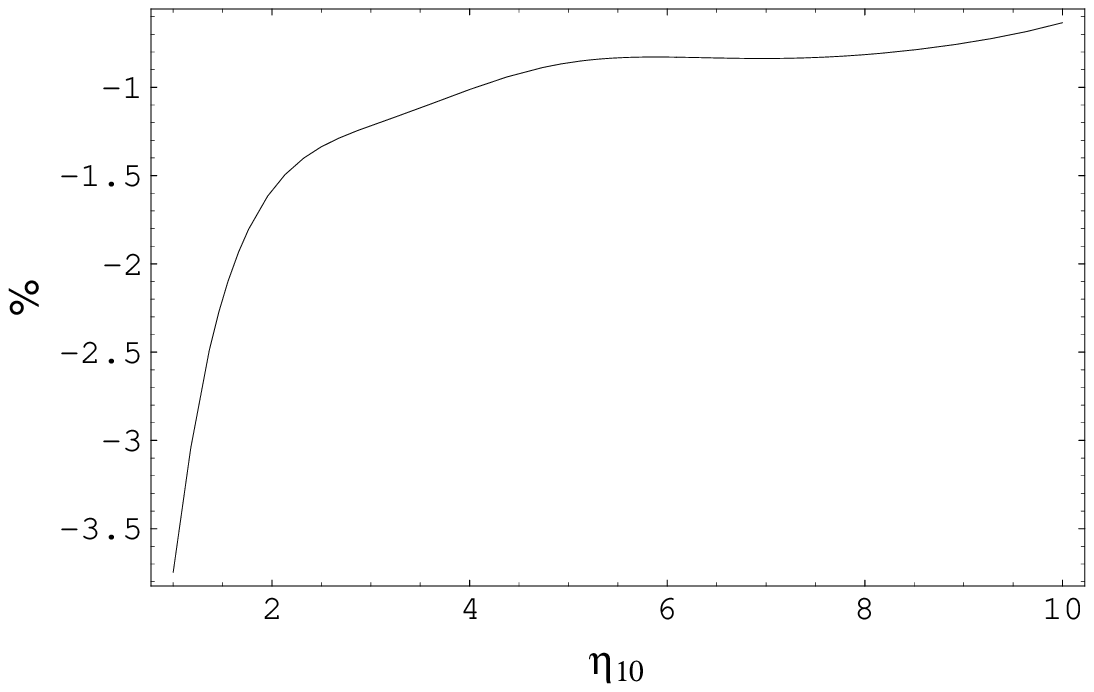,height=8cm}
\caption{Relative difference (in percent) for $Y_{^3He}$ between the
results of present work and the ones from \protect\cite{NuclearSigma} in
the case $N_\nu=3$.}
\label{f:comphe3}
\end{figure}

\begin{figure}
\epsfig{file=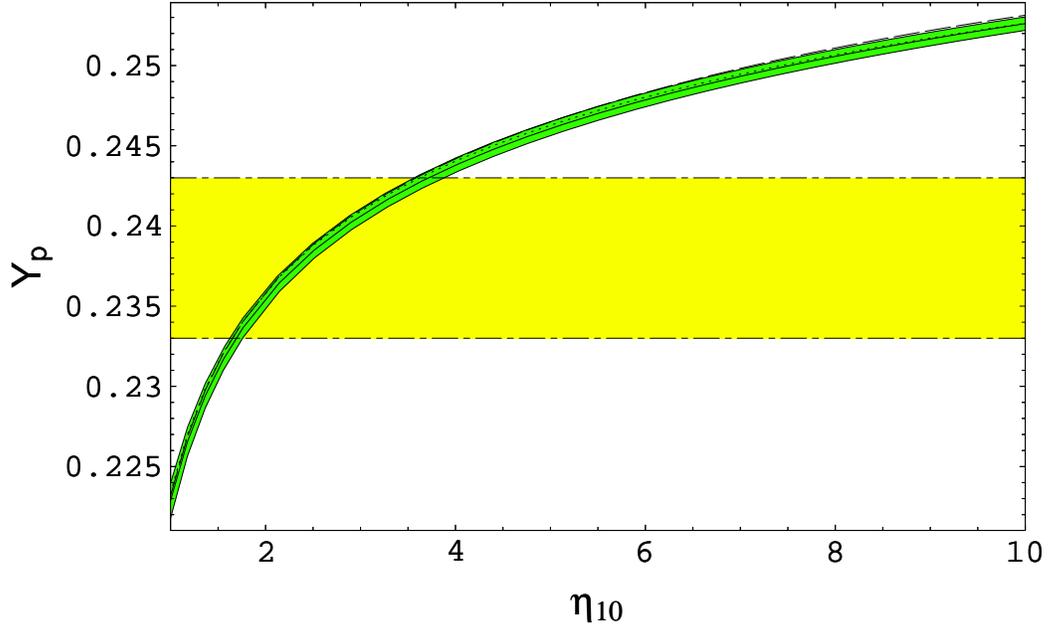,height=9cm}
\caption{The helium-4 mass fraction $Y_{p}$ \protect\eqn{e:abund4}, for
$N_\nu=3$, versus $\eta_{10}$. The dashed line is the result from
\protect\cite{Lopez} and the dotted one from \protect\cite{NuclearSigma}.}
\label{f:He4}
\end{figure}

\begin{figure}
\epsfig{file=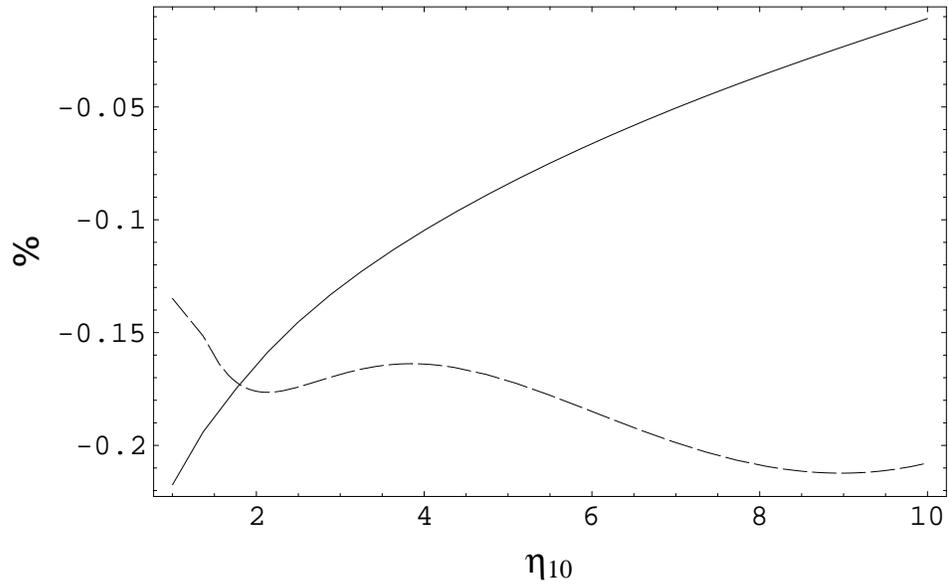,height=8.5cm}
\caption{Relative difference (in percent) for $Y_p$ between the results of
present work and the ones from \protect\cite{NuclearSigma} (solid line)
and from \protect\cite{Lopez} (dashed line) in the case $N_\nu=3$.}
\label{f:comphe4}
\end{figure}

\begin{figure}
\epsfig{file=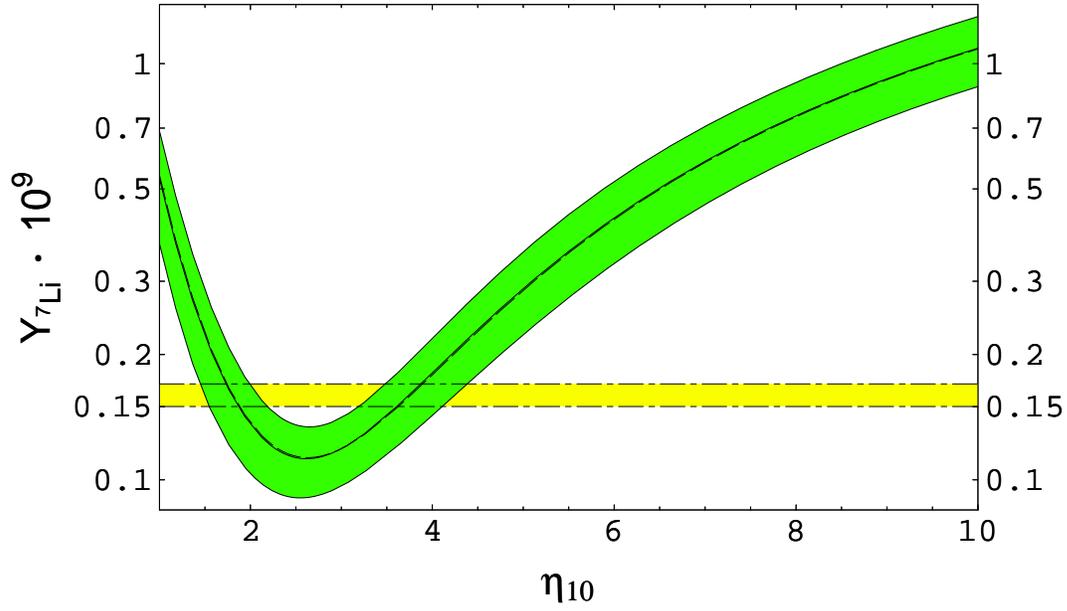,height=9cm}
\caption{The lithium-7 abundance $Y_{^7Li}$ \protect\eqn{e:abund237}, for
$N_\nu=3$, versus $\eta_{10}$. The dashed line, on this scale
indistinguishable from the central solid line, is the result from
\protect\cite{NuclearSigma}. No result is given for $^7Li$ in
\protect\cite{Lopez}.}
\label{f:Li7}
\end{figure}

\begin{figure}
\epsfig{file=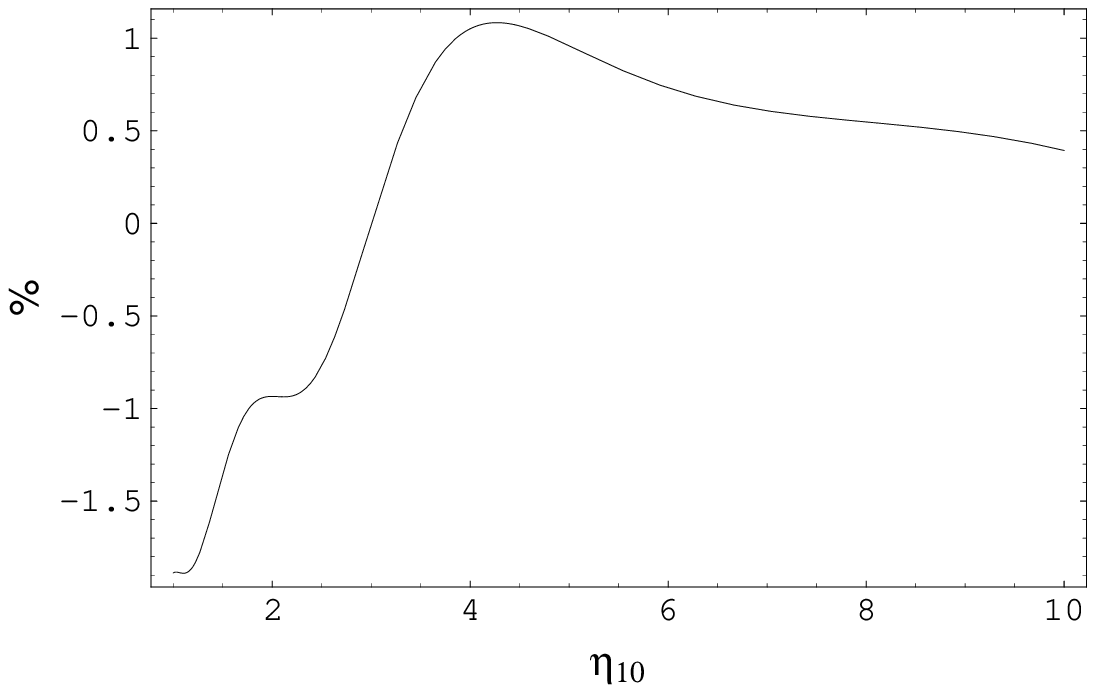,height=8cm}
\caption{Relative difference (in percent) for $Y_{^7Li}$ between the
results of present work and the ones from \protect\cite{NuclearSigma} in
the case $N_\nu=3$.}
\label{f:compli7}
\end{figure}

\begin{table}[t]
\begin{center}
\begin{tabular}{|c|c|c|c|c|c|}
\hline\hline
& $Y_D$ & $Y_{^3He}$ & $Y_p$ & $Y_p^m$ & $Y_{^7Li}$ \\
\hline\hline &&&&& \\
present & $0.3609{\cdot} 10^{-4}$ & $0.1166{\cdot} 10^{-4}$ & 0.2461 &
0.2448 & $0.2879{\cdot} 10^{-9}$ \\  analysis&&&&& \\
\hline &&&&& \\
previous & $0.3638{\cdot} 10^{-4}$ & $0.1175{\cdot} 10^{-4}$ & 0.2460 &
0.2447 & $0.2814{\cdot} 10^{-9}$ \\ analysis&&&&& \\
\hline&&&&& \\
\hline\hline
\end{tabular}
\end{center}
\caption{The predictions on light element abundances obtained with the new
BBN code, compared with our previous results \protect\cite{EMMP2}.}
\label{t:abund}
\end{table}
In Figures \ref{f:D}, \ref{f:He3}, \ref{f:He4}, \ref{f:Li7} we show the
theoretical predictions for the abundances \eqn{e:abund237}-\eqn{e:abund4},
compared with the experimental values. Figures \ref{f:compd},
\ref{f:comphe3}, \ref{f:comphe4}, \ref{f:compli7} show the comparison
between our results and the ones of Ref.s \cite{NuclearSigma,Lopez}. The
relative differences among our predictions and the results of Ref.s
\cite{NuclearSigma,Lopez} for $^4He$, in the relevant range for $\eta_{10}
\equiv  10^{10} \eta$, are less than $0.25 \%$, and thus probably due to
different ways of taking into account subdominant effects (thermal
radiative corrections). The differences are not much larger for Deuterium,
but reach few percents for $^3He$ and $^7Li$, which however have very large
theoretical errors due to the uncertainties on nuclear reaction rates. Note
that for $D$, $^3He$ and $^7Li$ only the results of \cite{NuclearSigma} are
available.
\begin{table}
\begin{center}
\begin{tabular}{|c|c|c|c|c|c|c|}
\hline\hline
$k_i\, {\cdot}\, Y_i$ & $j$   & $a_j$ & $b_j$ & $c_j$ & $d_j$ & $e_j$ \\
\hline\hline
$10^3 {\cdot} D$
  &  0  & 0.48212  & - & 0.076014  & 0.0015595 & -       \\
  &  1  & 0.12998  & - & 0.011704  & 0.0015438 & -4.0240 \\
  &  2  & 0.24279  & - & -0.056835 & -0.010037 & -1.1394 \\
  &  3  & -0.91776 & - & 0.0045225 & -0.014180 & 4.7712  \\
  &  4  & 2.2660   & - & 0.27257   & 0.090586  & -9.7616 \\
  &  5  & -1.1308  & - & -0.20773  & -0.15266  & 9.5377  \\
  &  6  & -1.4315  & - & -0.096193 & 0.11416   & -3.5489 \\
  &  7  & 1.0751   & - & 0.10664   & -0.031562 & -       \\
\hline
$10^5 {\cdot}~ ^3He$
  &  0  & 3.3201  & - & 0.18342  & -0.0067466 & -       \\
  &  1  & -8.8146 & - & -0.60698 & 0.013121   & 0.93721 \\
  &  2  & 7.3100  & - & 1.0607   & -0.042731  & 2.1792  \\
  &  3  & 14.543  & - & -0.52115 & 0.26524    & -5.6965 \\
  &  4  & -51.298 & - & -1.9291  & -0.72719   & 5.8729  \\
  &  5  & 76.092  & - & 4.5046   & 0.94031    & -8.2579 \\
  &  6  & -59.309 & - & -3.8725  & -0.58402   & 3.8572  \\
  &  7  & 20.591  & - & 1.2760   & 0.13864    & -       \\
\hline
$10 {\cdot}~ ^4He$
  &  0  & 2.2289    & 0.0020479   & 0.13021    & -0.0096485 & -        \\
  &  1  & -0.052824 & -0.00064726 & 0.0075762  & 0.0018598  & 0.27365  \\
  &  2  & 0.59320   & 0.0033813   & -0.013789  & -0.0030398 & -0.58807 \\
  &  3  & 0.49399   & -0.014272   & 0.082061   & -0.011842  & 0.22190  \\
  &  4  & 1.9579    & 0.041398    & 0.10938    & 0.027656   & -1.3763  \\
  &  5  & -0.89742  & -0.057323   & -0.12581   & -0.051066  & 0.93037  \\
  &  6  & 0.34269   & 0.040258    & 0.096350   & 0.039127   & -0.17224 \\
  &  7  & 0.47527   & -0.010753   & -0.0018617 & -0.013441  & -        \\
\hline
$10 {\cdot}~ ^4He_m$
  &  0  & 2.2184   & 0.0020421  & 0.13003   & -0.0095937 & -        \\
  &  1  & -0.44715 & -0.0011625 & -0.016604 & 0.0040428  & 0.44707  \\
  &  2  & 0.94939  & 0.0065040  & 0.010744  & -0.013907  & -0.69814 \\
  &  3  & 6.8365   & -0.026617  & 0.43785   & 0.018033   & -2.8141  \\
  &  4  & -1.1513  & 0.094984   & 0.0086552 & -0.12409   & -0.80120 \\
  &  5  & 2.0116   & -0.14739   & -0.17116  & 0.19005    & 3.0861   \\
  &  6  & 3.1081   & 0.11736    & 0.48418   & -0.16606   & -1.1642  \\
  &  7  & 4.0381   & -0.031739  & 0.088568  & 0.031865   & -        \\
\hline
$10^9 {\cdot}~ ^7Li$
  &  0  & 0.52920 & - & 0.15387   & 0.011486   & -       \\
  &  1  & -1.5617 & - & -0.42097  & -0.0075472 & -2.0190 \\
  &  2  & 2.0002  & - & -0.23058  & -0.21878   & -1.2259 \\
  &  3  & 0.58298 & - & 2.9150    & 0.94226    & 9.8088  \\
  &  4  & -8.3304 & - & -6.7141   & -1.7403    & -9.7328 \\
  &  5  & 22.627  & - & 6.2727    & 1.8244     & 3.1404  \\
  &  6  & -21.053 & - & -1.9273   & -1.1338    & 0.15166 \\
  &  7  & 6.1684  & - & -0.090045 & 0.32046    & -       \\
\hline\hline
\end{tabular}
\end{center}
\bigskip
\caption{Values of the coefficients of Eq.~\protect\eqn{e:yfit} for light
element abundances.}
\label{t:fitcoef}
\end{table}

We have performed a fit of the previous abundances as functions of $N_\nu$,
$\tau_n$, and $x \equiv \log_{10} \eta_{10}$, with an accuracy which is
better than $1\%$ in the all ranges $0.5\leq N_\nu\leq 6$, $882.9\, s \leq
\tau_n \leq 890.5\, s$, $0\leq x\leq 1$. In particular, for $N_\nu
= 3$ and $^4He$, our fit is accurate at $0.01\%$. The fitting functions have
been chosen as
\bea
k_i\, {\cdot}\, Y_i\, (N_\nu,~ \tau_n,~ x) &=& \left[~ \sum_{j=0}^8 a_j\,
x^j + \left(\, \sum_{j=0}^8 b_j\, x^j\, \right)\, ( \tau_n - \tau^{ex}_n )
+ \left(\, \sum_{j=0}^8 c_j\, x^j\, \right)\, ( N_\nu - 3 )~ + \right. \nn
\\
&& \left(\, \sum_{j=0}^8 d_j\, x^j\, \right)\, ( N_\nu - 3 )^2~ \biggr]
\exp \left\{~ \sum_{j=1}^6 e_j\, x^j~ \right\} \vv
\label{e:yfit}
\eea
with all coefficients given in Table \ref{t:fitcoef}.

By using these expressions \eqn{e:yfit}, as the theoretical predictions for
the light element abundances, and their experimental measurements
\eqn{e:olive99}, \eqn{e:lowD}, \eqn{e:highD}, \eqn{e:Li7} it is possible to
test the compatibility of standard BBN scenario in the $N_\nu$--$\eta_{10}$
plane. To this end, we define a {\it total likelihood function} as
\be
{\cal L} (N_\nu,\, \eta_{10}) = L_D (N_\nu,\, \eta_{10})~ L_{^4He}
(N_\nu,\, \eta_{10})~ L_{^7Li} (N_\nu,\, \eta_{10}) \vv
\label{e:like}
\ee
where the likelihood function for each abundance, assuming Gaussian
distribution for the errors, is given by the overlap
\bea
&L_i (N_\nu,\, \eta_{10}) =& \nn \\
& \!\!\!\!\!\!\!\!\!\!\! = \displaystyle \frac{1}{2\, \pi\, \sigma_i^{th}
(N_\nu,\, \eta_{10})\, \sigma_i^{ex}}& \int dY~ \exp \left\{ -
\frac{(Y-Y_i^{th} (N_\nu,\, \eta_{10}))^2}{2\, \sigma_i^{th\, 2} (N_\nu,\,
\eta_{10})} \right\} \exp \left\{ - \frac{(Y-Y_i^{ex})^2}{2\,
\sigma_i^{ex\, 2}} \right\}
\label{e:likei}
\eea
In order to evaluate $L_i (N_\nu,\, \eta_{10})$ we need the theoretical
uncertainties $\sigma_i^{th} (N_\nu,\, \eta_{10})$. In
Ref.~\cite{NuclearSigma} a new, alternative approach to the standard Monte
Carlo technique \cite{Montecarlo} has been proposed. This method is based
on linear error propagation for the estimate of theoretical uncertainties
on primordial abundances  due to the poor knowledge of  nuclear reaction
rates. The light element abundances, $Y_i$, and the logarithmic derivatives
of $Y_i$ with respect to the nuclear rates, $\lambda_{ik}$, are given as
polynomial fits, while the variation of the abundances for a change $\delta
R_k$ in the rate $R_k$ is obtained as
\be
\delta Y_i = Y_i~ \sum_k~ \lambda_{ik}\, \frac{\delta R_k}{R_k} \pp
\ee
Taking into account the error correlation, the error matrix results
\be
\sigma_{ij}^2 = Y_i~ Y_j~ \sum_k~ \lambda_{ik}\, \lambda_{jk}\, \left(
\frac{\Delta R_k}{R_k} \right)^2 \vv
\ee
where $\Delta R_k$ are the $1\sigma$ uncertainties. In particular the
theoretical $1\sigma$ uncertainties are given by the square root of the
diagonal elements,
\be
\sigma_i^{th} = \sqrt{\sigma_{ii}^2} \pp
\ee
We used the Fortran code provided by the authors \cite{Fastbbn} to
calculate the theoretical uncertainties $\sigma_i^{th} (N_\nu,\,
\eta_{10})$, which we used in Eq.~\eqn{e:likei}.

\begin{figure}
\centerline{\epsfig{file=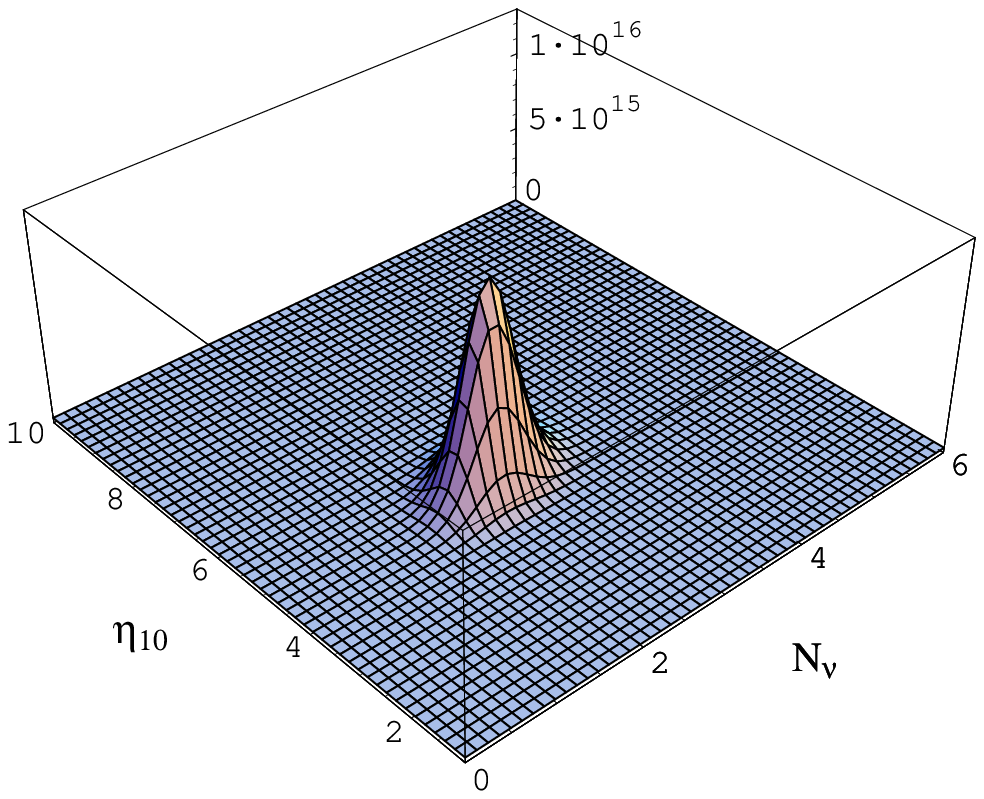,height=9cm}}
\caption{Total likelihood function \protect\eqn{e:like} versus
$N_\nu-\eta_{10}$ for the low experimental value of $D$
\protect\cite{Songaila}.}
\label{f:likelow}
\end{figure}

\begin{figure}
\centerline{\epsfig{file=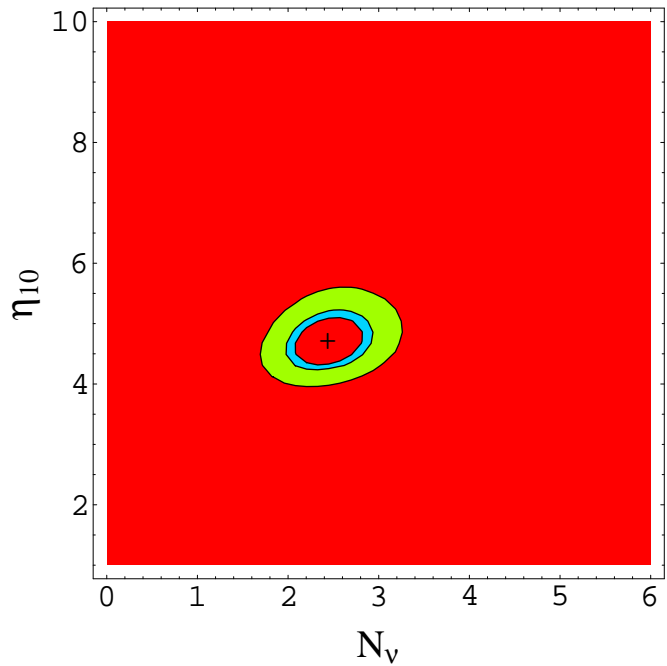,height=9cm}}
\caption{Contour plots of the total likelihood function for the low
experimental value of $D$. From inner to outer they correspond to $50\%$,
$68\%$ and $95\%$ CL, respectively. The cross indicates the maximum of
likelihood function, and corresponds to $N_\nu = 2.44$ and $\eta_{10}=
4.69$.}
\label{f:conlow}
\end{figure}

\begin{figure}
\centerline{\epsfig{file=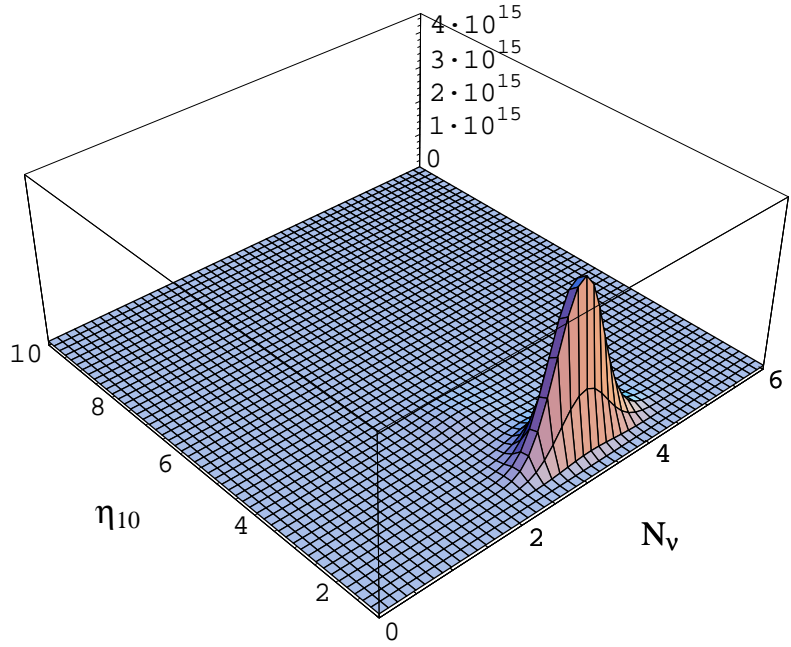,height=9cm}}
\caption{Total likelihood function versus $N_\nu-\eta_{10}$ for the high
experimental value of $D$ \protect\cite{Tytler}.}
\label{f:likehigh}
\end{figure}

\begin{figure}
\centerline{\epsfig{file=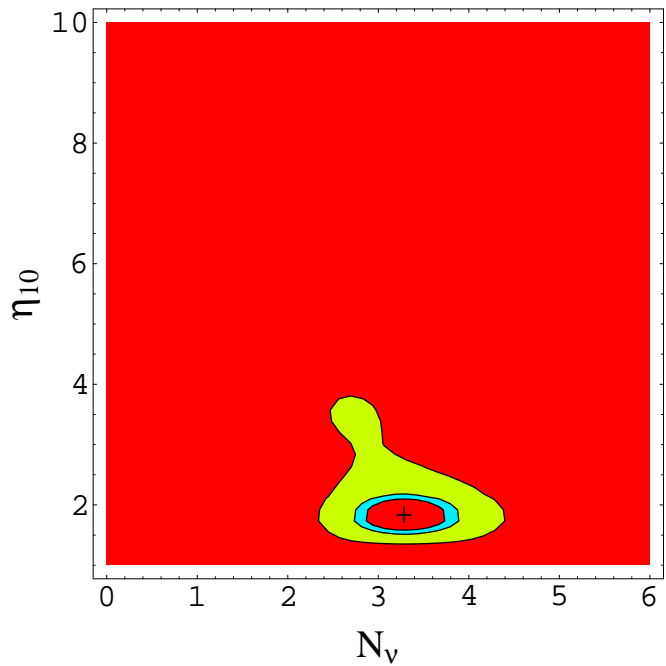,height=9cm}}
\caption{Contour plots of the total likelihood function for the high
experimental value of $D$. From inner to outer they correspond to $50\%$,
$68\%$ and $95\%$ CL, respectively. The cross corresponds to $N_\nu = 3.29$
and $\eta_{10}= 1.81$.}
\label{f:conhigh}
\end{figure}

The total likelihood function \eqn{e:like} and the corresponding contour
plots for $50\%$, $68\%$ and $95\%$ CL, for low and high $D$ are shown in
Figures~\ref{f:likelow}, \ref{f:conlow} and \ref{f:likehigh},
\ref{f:conhigh}, respectively. As already clear from Figure \ref{f:D}, the
two different experimental estimates of $D$ single out different regions
for $\eta_{10}$. From the $95\%$ CL contour of Figure \ref{f:conlow}, for
low $D$ we have $4.0 \leq \eta_{10} \leq 5.7$, which is in fair agreement
with the similar results of Ref. \cite{olive99}, whereas for high $D$ from
Figure \ref{f:conhigh} we get $1.4 \leq \eta_{10} \leq 3.7$.

\begin{figure}
\centerline{\epsfig{file=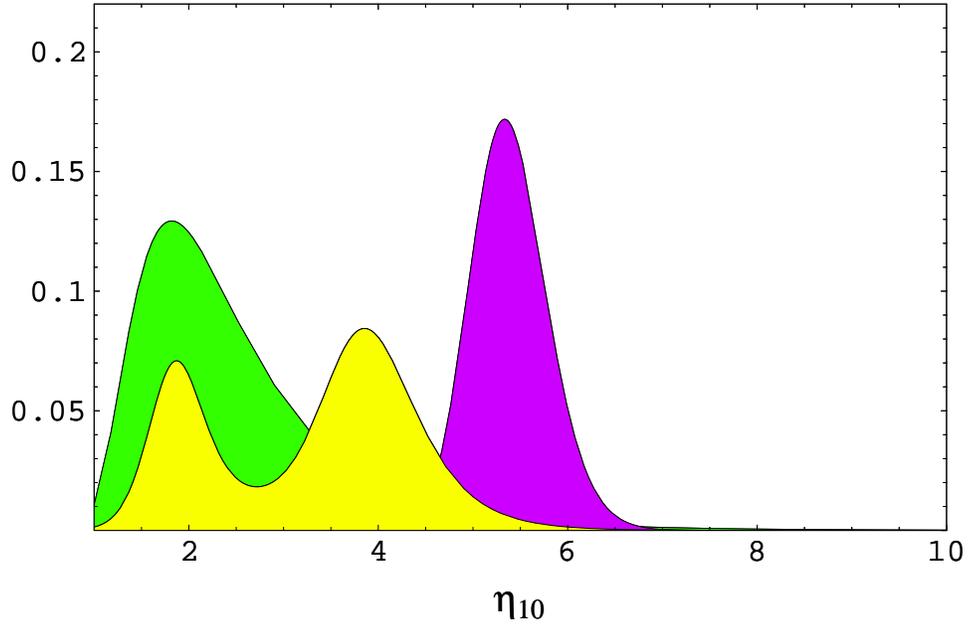,height=9cm}}
\caption{Single likelihood functions for the three light abundances in the
case of the low experimental value of $D$ and $N_\nu=3.29$. The single
peaked curve stands for $L_D$, the double peaked curve for $L_{^7Li}$ and
the broad one for $L_{^4He}$. The same notation holds for the following
figures.}
\label{f:3likelow}
\end{figure}

\begin{figure}
\centerline{\epsfig{file=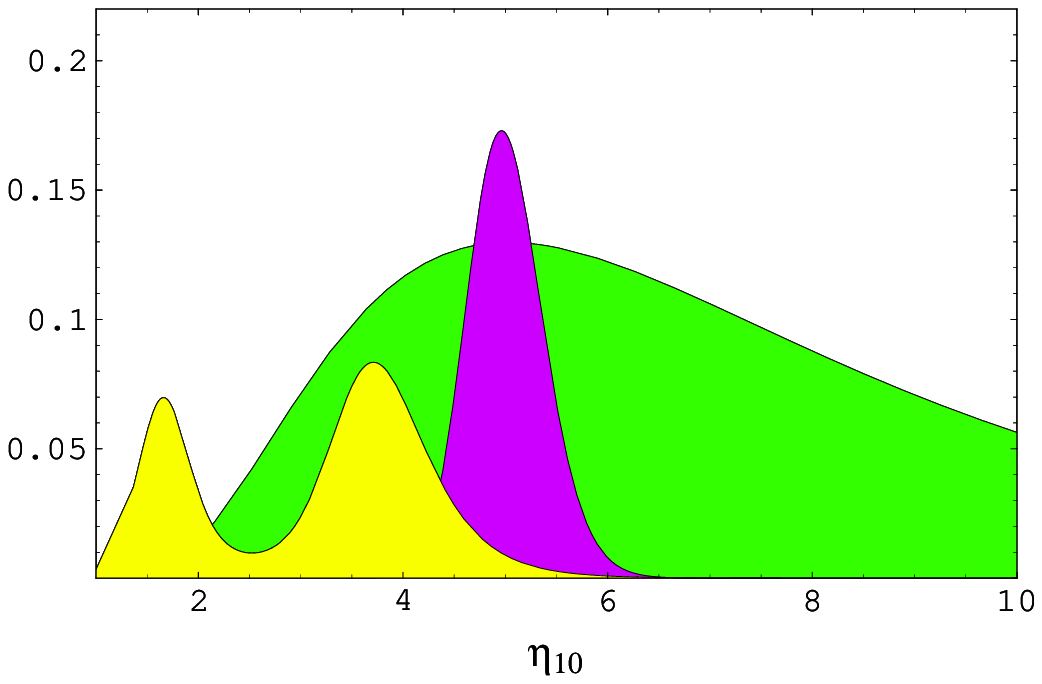,height=9cm}}
\caption{Single likelihood functions for the three light abundances in the
case of the low experimental value of $D$ and $N_\nu=2.44$.}
\label{f:2likelow}
\end{figure}

\begin{figure}
\centerline{\epsfig{file=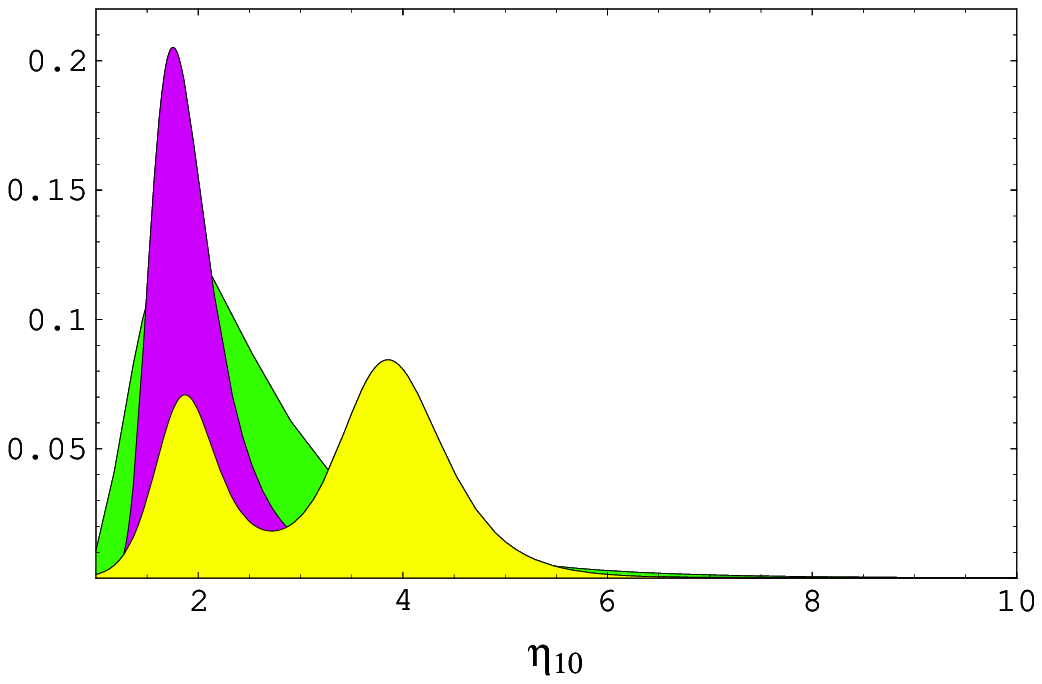,height=9cm}}
\caption{Single likelihood functions for the three light abundances in the
case of the high experimental value of $D$ and $N_\nu=3.29$.}
\label{f:3likehigh}
\end{figure}

\begin{figure}
\centerline{\epsfig{file=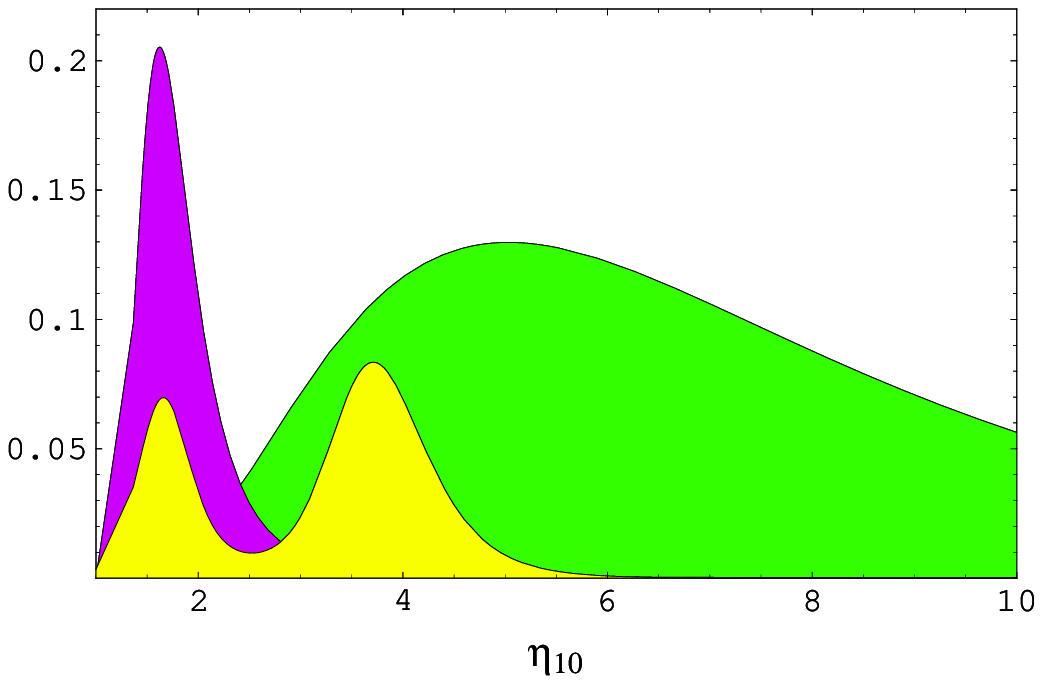,height=9cm}}
\caption{Single likelihood functions for the three light abundances in the
case of the high experimental value of $D$ and $N_\nu=2.44$.}
\label{f:2likehigh}
\end{figure}

As far as $N_\nu$ is concerned, for low $D$ and for $95 \%$ CL one has $1.7
\leq N_\nu \leq 3.3$, whereas for high $D$ one gets $2.3 \leq N_\nu \leq
4.4$. In both cases we obtain comparable ranges for $N_\nu$ with respect to
Ref. \cite{olive99}. The total likelihood function is peaked around the
points shown as the crosses in Figures \ref{f:conlow} and \ref{f:conhigh},
which correspond to $N_\nu = 2.44$ and $\eta_{10}=4.69$ for low $D$, and
to $N_\nu = 3.29$ and $\eta_{10}=1.81$ for high $D$. The position of the
two maxima is easily understood. The Deuterium abundance is a decreasing
function of $\eta_{10}$. Since lowering $\eta_{10}$ results in a smaller
$^4He$ mass fraction $Y_p$, it is necessary to compensate this effect by
increasing the universe expansion rate via a larger $N_\nu$. This in fact
leads to a larger value for the freeze--out temperature for nucleon weak
interactions and thus gives a larger amount of the initial neutron to
proton density ratio. The single contributions \eqn{e:likei} to the total
likelihood \eqn{e:like} can be easily recognized by looking at Figures
\ref{f:3likelow}-\ref{f:2likehigh}, where we show the $D$, $^4He$ and
$^7Li$ likelihood functions, for both high and low Deuterium results, for
the two preferred values $N_\nu=2.44$ and $N_\nu=3.29$. For the low $D$
case, a better overlap between the maxima of the single likelihoods of $D$
and $^4He$ is realized for $N_\nu=2.44$. The opposite situation occurs for
high $D$ where to $N_\nu=3.29$ corresponds a better overlap of the single
likelihoods.

\section{BBN predictions for degenerate neutrinos}

The assumption of vanishing neutrino chemical potentials can be only
justified by the sake of simplicity or by the theoretical prejudice that
their order of magnitude should be set by the ones of the corresponding
charged leptonic partner. However, physical scenarios in which large lepton
asymmetries are produced, which do not lead to a large baryon asymmetry,
have been proposed in literature \cite{AF,Casas,MMR,Mac}. They are based on
the Affleck-Dine mechanism \cite{AF} or on active--sterile neutrino
oscillations \cite{Foot,Pastor}. In particular, the expected asymmetry may
be different for each neutrino family. For these reasons, at least in
principle, it is worth-while considering the primordial nucleosynthesis in
presence of large neutrino asymmetries, i.e. for non-vanishing neutrino
chemical potentials. It seems to us that this topic, which has been
extensively studied in past \cite{KangSteigman}, receives a renewed
interest in view of the recent BOOMERANG and MAXIMA-1 results
\cite{Boomerang} on the acoustic peaks of the CMBR, which suggests a larger
value of the baryonic matter contribution to the total energy density
$\Omega$. In their analysis of the data they find a baryon density
$\Omega_b h^2 \sim 0.02 \div 0.03$, or $\eta \sim (6 \div 10) 10^{-10}$,
which, as is clear from the considerations of the previous section, is
incompatible at $95 \%$ CL with the standard BBN result (see Figures
\ref{f:conlow} and \ref{f:conhigh}). Actually even larger values for
$\Omega_b h^2$ are obtained if no constraints are imposed in the
likelihood analysis (see \cite{Boomerang}). It is the aim of this section
to discuss whether a finite neutrino chemical potential may reconcile BBN
theoretical predictions, the observed nuclide abundances and a higher
value for $\eta$. In particular we report the results of a new analysis of
the degenerate BBN scenario we have performed with our code.

The effect of neutrino chemical potentials on BBN predictions is twofold.
Due to the definition of $N_\nu$ \eqn{neff}, non-vanishing
$\xi_\alpha=\mu_{\nu_\alpha}/T_\nu$, with $\alpha$ denoting the neutrino
specie, change its value from the non-degenerate case ($\xi_\alpha=0$). In
fact for three massless neutrinos with degeneracy parameter $\xi_\alpha$,
$N_\nu$ becomes
\be
N_{{\nu}} = 3 + \Sigma_{\alpha =e,\mu, \tau} \left[ \frac{30}{7} \left(
\frac{\xi_\alpha}{\pi} \right)^2 + \frac{15}{7} \left(
\frac{\xi_\alpha}{\pi} \right)^4 \right] \vv
\label{neff1}
\ee
implying a larger expansion rate of the universe with respect to the
non-degenerate scenario. In this case, nucleons freeze out at a larger
temperature, with a higher value for the neutron to proton density ratio,
which implies a larger value of $Y_p$. This effect does not depend on the
particular neutrino chemical potentials but rather on the whole
neutrino--antineutrino asymmetry, via the sum in the r.h.s. of
\eqn{neff1}. In addition, we also note that the neutrino decoupling
temperature and the ratio $T_\nu / T$, entering in the BBN equations, are
affected by a change of $N_\nu$ as well. However, it has been checked
\cite{Trieste} that this effect is quite negligible on the predictions for
the element abundances.

Electron neutrinos entering in the $n \leftrightarrow p$ processes, can
modify the corresponding rates if their distribution has a non-vanishing
$\xi_e$. In particular, a positive value for $\xi_e$ means a larger number
of $\nu_e$ with respect to $\bar{\nu}_e$ and thus enhances $n \rightarrow
p$ processes with respect to the inverse processes. This, of course,
reduces the number of neutrons available at the onset of BBN. Moreover,
since the initial condition for BBN at $T \sim 10~MeV$ is fixed by the
nuclear thermal equilibrium, which is also kept by reactions like $\nu_e +
n \rightarrow p + e^-$, the chemical equilibrium fixes $(\mu_p - \mu_n)/T
\approx \xi_e$. This implies an initial $n/p$ ratio which is lowered by the
factor $\exp(-\xi_e)$, which again reduces the number of neutrons available
at the onset of BBN.

These effects strongly influence the nuclide production, so with no
$\nu_\mu$ and $\nu_\tau$ degeneracy, the value of $\xi_e$ is strongly
constrained. In this  case, the authors of \cite{KangSteigman} found the
limits $-0.06 \leq \xi_e \leq 0.14$. For a fully degenerate BBN, at least
for $Y_p$, the effect of a positive $\xi_e$ can be compensated by the
contribution to $N_\nu$ coming from $\xi_{\mu,\tau}$. In
\cite{KangSteigman} the neutrino degeneracy parameters result to be in the
ranges $-0.06 \leq \xi_e \leq 1.1$ and $|\xi_{\mu, \tau}| \leq 6.9$ for
$\xi_\mu \neq 0$, $\xi_\tau =0$ or viceversa, and $|\xi_{\mu, \tau}| \leq
5.6$ for $\xi_\mu = \xi_\tau \neq 0$.

We do consider in our analysis as input parameters $\xi_e$, $N_\nu$ and
$\eta_{10}$. Likelihood analysis of compatibility between theoretical
predictions and experimental data yields contour levels which are surfaces
in the three-dimensional space of these parameters. Because of the partial
cancellation of the effects due to $\xi_e$ and $\xi_{\mu,\tau}$ we have
discussed, if we define, in analogy to the non--degenerate case, a total
likelihood function, ${\cal L} (\xi_e,\, N_\nu, \,\eta_{10})$, it is
reasonable to expect that it may sensibly differ from zero in a quite wide
parameter region, so some bound should be chosen to the possible range of
variation for the parameters. We have chosen to constrain $N_\nu$ to be
smaller than 13. This bound has been obtained, at $2\sigma$ level, in
\cite{Hannestad}, by a likelihood analysis of the BOOMERANG data, as
function of $N_\nu$, maximizing for each $N_\nu$ the likelihood function
over all other parameters, including $\eta$. As we will see our conclusion
on the degenerate BBN scenario versus BOOMERANG and MAXIMA-1 data is
completely different than what has been argued in $\cite{Hannestad}$.
Nevertheless we think that the upper bound on $N_\nu$ is quite robust. We
also consider $\xi_e$ and $\eta_{10}$ in the wide range $-1 \div 1$ and
$1\div 30$, respectively.

By using the results of our BBN code for the degenerate neutrino case and
performing a study similar to the one presented for the standard scenario,
we obtained likelihood functions for both low and high $D$ experimental
values. We first report the maxima for these functions in the low $D$ case
\be
\xi_e = 0.06 \vv \qquad N_\nu = 3.43 \vv \qquad \eta_{10} = 5 \vv
\label{maxdl}
\ee
and in the high $D$ scenario
\be
\xi_e = 0.35 \vv \qquad N_\nu = 13 \vv \qquad \eta_{10} = 4.20 \vv
\label{maxdh}
\ee
We notice that in the low Deuterium case the solution prefers an almost
non degenerate scenario, with values for $\eta$ compatible with the one
obtained in the previous section and a slightly larger result for $N_\nu$.

\begin{figure}
\centerline{\epsfig{file=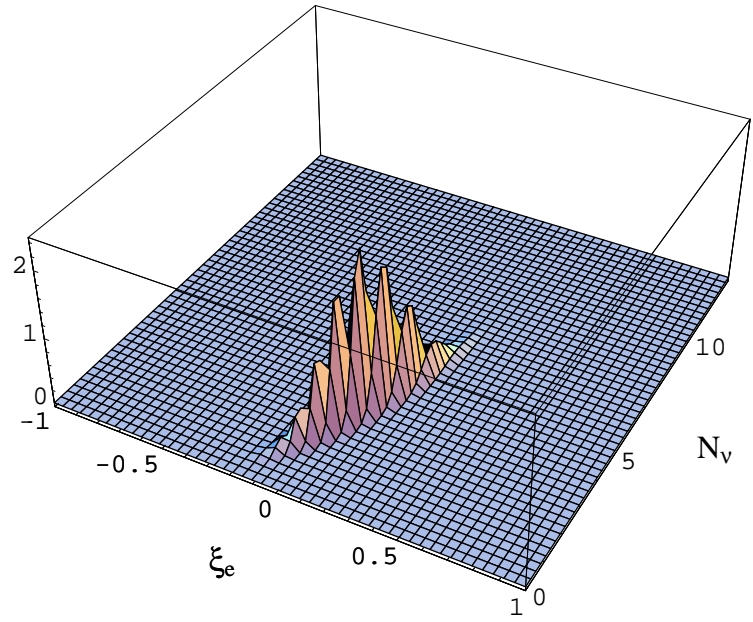,height=9cm}}
\caption{Total likelihood function in the degenerate BBN scenario in the
plane $\xi_e$-$N_\nu$, in the case of the low experimental value of $D$ for
$\eta_{10}=5$.}
\label{f:likedl}
\end{figure}

\begin{figure}
\centerline{\epsfig{file=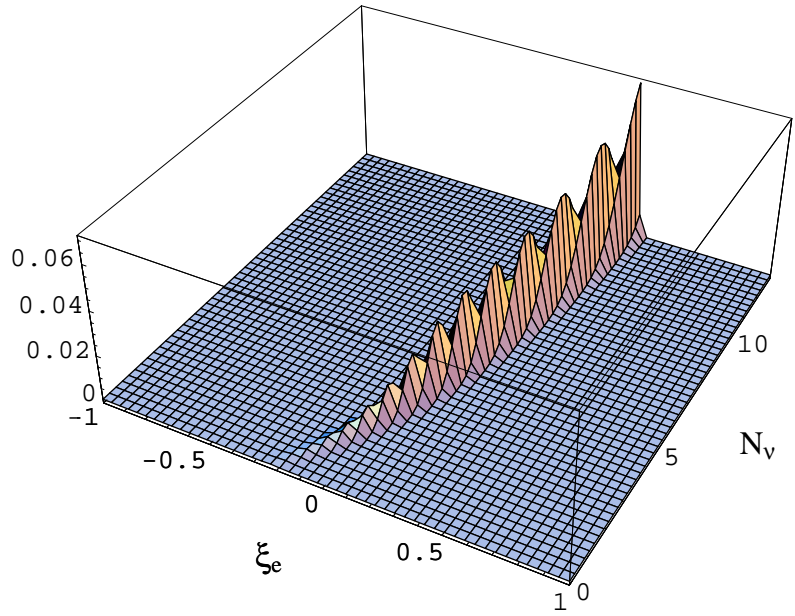,height=9cm}}
\caption{Total likelihood function in the degenerate BBN scenario, in the
case of the high experimental value of $D$ for $\eta_{10}=4.20$.}
\label{f:likedh}
\end{figure}

\begin{figure}
\epsfig{file=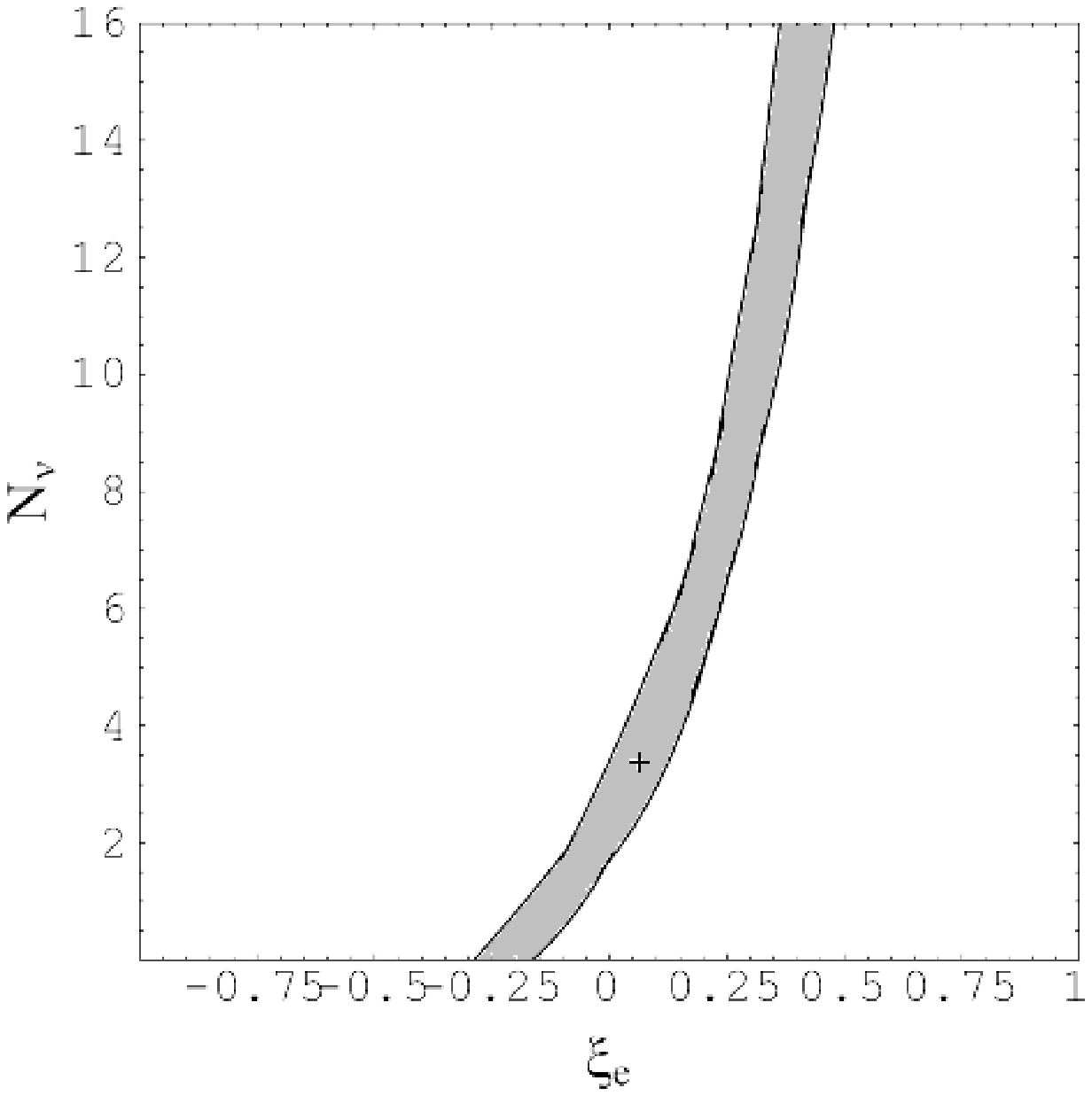,height=13cm}
\caption{The $95\%$ exclusion plot for the variables $\xi_e-N_\nu$ for low
$D$ experimental value. The dark area represents the BBN allowed region,
for $\eta_{10}$ in the range $3.3 \div 9.9$. The cross corresponds to the
maximum of the likelihood function.}
\label{f:condlow}
\end{figure}

\begin{figure}
\epsfig{file=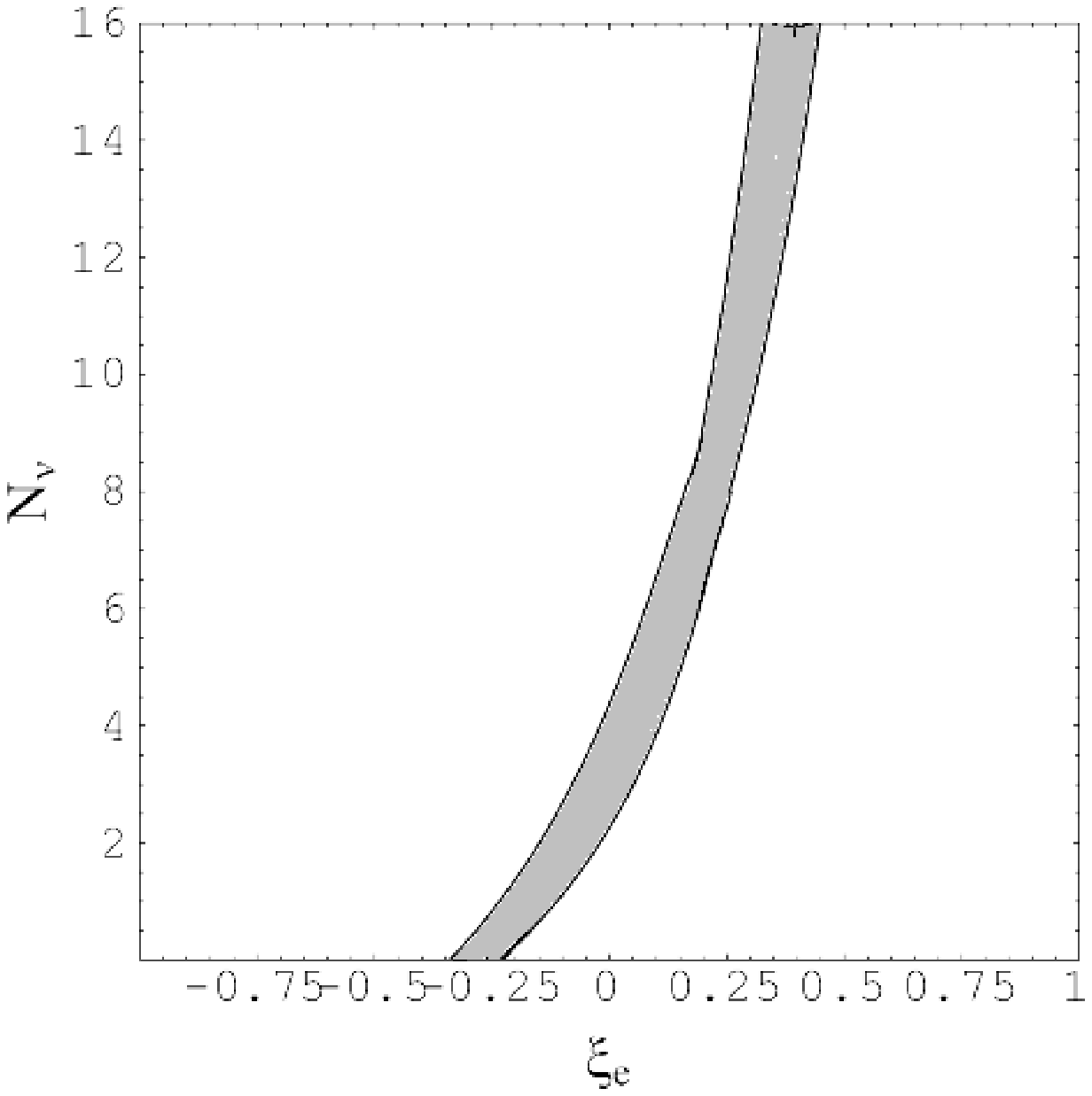,height=13cm}
\caption{The $95\%$ exclusion plot for the variables $\xi_e-N_\nu$ for
high $D$ experimental value. The dark area represents the BBN allowed
region, for $\eta_{10}$ in the range $1.1 \div 5.8$. The cross corresponds
to the maximum of the likelihood function.}
\label{f:condhigh}
\end{figure}

More interesting is to follows the allowed ranges for $\xi_e$ and $N_\nu$
as functions of $\eta$. In Figures \ref{f:likedl} and \ref{f:likedh} we
show ${\cal L} (\xi_e, \, N_\nu, \, \eta_{10})$, evaluated for $\eta_{10}$
taking the values in \eqn{maxdl} and \eqn{maxdh}. Increasing $\eta_{10}$
in the considered range, the likelihood functions {\it move} towards
higher values of $\xi_e$ and $N_\nu$, along, approximatively, a linear
path. This is due to the fact that increasing $\xi_e$, which results in a
lower value for the $n/p$ ratio, and then for $Y_p$, must be compensated
by a faster expansion produced by a higher value for $N_\nu$. This
behaviour is highlighted in the $95 \%$ exclusion plots for the $\xi_e$
and $N_\nu$ parameters for different values of $\eta_{10}$, which are
reported in Figures \ref{f:condlow} and \ref{f:condhigh} with the bound
$N_\nu \leq 13$. These two plots summarize our analysis of the degenerate
BBN scenario, providing the combined bounds on $\xi_e$ and $N_\nu$. For
low $D$ the allowed range for $\eta_{10}$ is $3.3 \div 9.9$, while for
high $D$ we have $1.1 \div 5.8$. In both cases the degenerate BBN
scenario, at $2 \sigma$ level is compatible with observational results on
nuclide abundances even for quite large values for both $N_\nu$ and
$\xi_e$. As expected, however, these two parameters are strongly
correlated.

From our analysis we see that a large value for $\eta \sim 10^{-9}$ is
compatible, at $2 \sigma$ level, with degenerate BBN for quite large
$N_\nu$ only, $N_\nu \geq 10$, and large positive $\xi_e \geq 0.3$. The
BOOMERANG and MAXIMA-1 results seem to indicate a value for $\eta \sim
10^{-9}$, but at the same time suggest $N_{\nu}\leq 13$, again at $2
\sigma$ level \cite{Hannestad}. We may conclude that the degenerate BBN
scenario and BOOMERANG and MAXIMA-1 results are compatible only in the very
top area in the exclusion plot of Figure (\ref{f:condlow}) for the low $D$
scenario. In the high $D$ case, the upper limit on $N_\nu$ is already
reached for $\eta \sim 6.5 {\cdot} 10^{-10}$. As a conclusion we can say that, if
the result $\eta \sim 10^{-9}$ will be confirmed, the degenerate BBN
scenario is not consistent with CMBR results at $2 \sigma$ level, while
there is agreement for the low $D$ case. For slightly lower values for
$\eta \sim ( 0.5 \div 0.6 ) 10^{-10}$, which corresponds to the central
value measured by MAXIMA-1 experiment, the agreement at $2 \sigma$ level
definitely improves, and it would represent a signal in favour of a large
neutrino--antineutrino asymmetry in the universe, $0.2 \leq \xi_e
\leq 0.5$ and $N_\nu \geq 10$ \cite{LP}. We note that a similar result has
been obtained in \cite{japan}.

\section{Conclusions}

In this paper we have reported the results of a likelihood analysis of Big
Bang Nucleosynthesis theoretical predictions versus the experimental data,
both for vanishing neutrino chemical potentials (standard BBN) and in
presence of neutrino-antineutrino asymmetries (degenerate BBN). The
theoretical estimates have been obtained by using a new updated code we
have developed in the recent years to increase the accuracy at the $1 \%$
level.

In the standard scenario, BBN predictions are in good agreement with
experimental data for both low and high $D$ results. In the first case,
since $Y_D$ is a decreasing function of baryonic asymmetry $\eta$, the low
Deuterium abundance fixes the maximum of total likelihood to larger values
of $\eta$, and to compensate the effect on $Y_p$ yields smaller values of
$N_\nu$. In this way we have obtained as preferred values $N_\nu= 2.44$ and
$\eta_{10}=4.69$. The situation is reversed for high $D$ where the maximum
lies at $N_\nu = 3.29$ and $\eta_{10}=1.81$. From the plots of $95\%$ CL
for both the above likelihood functions one gets the corresponding
compatibility regions
\bea
\begin{array}{ccccc}\mbox{low}~D  \,\,\, & 1.7 \leq N_\nu \leq 3.3 & & 4.0
\leq \eta_{10} \leq 5.7 &  \,\,\, 0.015 \leq \Omega_B h^2 \leq 0.021 \\
&&&& \\ \mbox{high}~D  \,\,\, & 2.3
\leq N_\nu \leq 4.4 & & 1.4 \leq \eta_{10} \leq 3.7 &  \,\,\, 0.005 \leq
\Omega_B h^2 \leq 0.014
\end{array} \pp
\label{ranges}
\eea
Both ranges on $N_\nu$ are comparable with the results of Ref.
\cite{olive99}. For the baryon asymmetry $\eta$, the result for low $D$
fairly coincides with \cite{olive99}, whereas the distortion of the contour
for $95\%$ CL in case of high $D$ makes our upper limit on $\eta_{10}$
larger than the value of \cite{olive99}. This effect is mainly due to the
non-trivial dependence on the parameters $N_\nu-\eta_{10}$ of the nuclear
theoretical uncertainties.

We have also analyzed the degenerate BBN scenario, motivated by the
BOOMERANG and MAXIMA-1 results \cite{Boomerang} on the CMBR spectrum. For
both high and low Deuterium measurements we have obtained a $2 \sigma$
exclusion plots for the $\xi_e$ and $N_\nu$ parameters, and we have shown
that the theoretical estimates are in good agreement with the experimental
measurements of light nuclei abundances even for large values of these
parameters, provided they lie in the regions shown in Fig.s
(\ref{f:condlow}) and (\ref{f:condhigh}). In particular large values for
$\eta$ implies high values for the neutrino degeneracy parameters.

A first analysis of the recent CMBR data, \cite{Boomerang}, suggests quite
large values for $\eta \sim 10^{-9}$. From our results we see that this
baryon to photon density is actually too high to be in good agreement with
degenerate BBN, the compatibility being only at $2 \sigma$ level, with the
low Deuterium scenario. This conclusion holds if we constraint the
effective number of neutrinos to be bounded by $N_\nu \leq 13$, which, as
pointed out in \cite {Hannestad}, is again suggested by BOOMERANG data.
The agreement improves for slightly smaller values for $\eta$, as the
central value obtained by MAXIMA-1 results, and would be an evidence in
favour of large neutrino degeneracy, $0.2 \leq \xi_e \leq 0.5$, $N_\nu
\geq 10$, an intriguing feature of the neutrino cosmic background. New
experimental results, as the one expected from the MAP and PLANCK
experiments on CMBR, as well as new analysis of the BOOMERANG and MAXIMA-1
data, will be crucial in clarifying this issue.\\

\noindent
{\large \bf Acknowledgements}\\ The authors are pleased to thank Dr.s
Sergio Pastor, Marco Peloso and Francesco Villante for useful discussions
and comments. We thank A. Bottino for pointing us the MAXIMA-1
Collaboration results, and M. Kaplinghat for valuable remarks.

\appendix
\section{Series expansion for integral quantities}
\setcounter{equation}0
The partial derivatives of $L$ with respect to $z$ and $\pe$, $L_z$ and
$L_{\pe}$, can be expressed as the following series\footnote{We truncate
all the series at $n=7$ while in the standard code \cite{Kawano} the
truncation is at $n=5$.}
\bea
L_z (z,\, \pe) = \frac{d z_R}{dz}~ \left\{ \frac{2\, L}{z_R} - \left(
\frac{z_R}{\pi} \right)^2 \sum_{n=1}^\infty~ (-1)^{n+1}~ \left[ K_1 (n\,
z_R) + K_3 (n\, z_R) \right]\, \sinh (n\, \pe) \right\}, \\
L_{\pe} (z,\, \pe) = \frac{2\, z_R^2}{\pi^2}~ \sum_{n=1}^\infty~
(-1)^{n+1}~ K_2 (n\, z_R)\, \cosh (n\, \pe) \vv ~~~~~~~~~~~~~~~~~
\eea
where the $K_i$ are the modified Bessel functions of the second kind and in
$z_R (z) = m_e^R/T$ we consider the renormalized electron mass,
\be
m_e^R (T) \simeq m_e~ \left[ 1 + \frac \pi 3\, \alpha~ \left( \frac{T}{m_e}
\right)^2 \right].
\ee
The series expansion for the remaining electron quantities are
\bea
\hrho_e (z,\, \pe) &=& \frac{z_R^3}{2\, \pi^2}~ \sum_{n=1}^\infty~
\frac{(-1)^{n+1}}{n}~ \left[ 3\, K_3 (n\, z_R) + K_1 (n\, z_R) \right]\,
\cosh (n\, \pe) \vv \\
\frac{\delta \hrho_e}{\delta z} (z,\, \pe) &=& \frac{d z_R}{dz}~ \biggl\{
\frac{3\, \hrho_e}{z_R} + \frac{z_R^3}{2\, \pi^2} \sum_{n=1}^\infty~
(-1)^{n+1}~ \biggl[ \frac{15\, K_3 (n\, z_R) + K_1 (n\, z_R)}{n\, z_R} \nn
\\
&& - 4\, K_4 (n\, z_R) \biggr]\, \cosh (n\, \pe) \biggr\} \vv \\
\frac{\delta \hrho_e}{\delta \pe} (z,\, \pe) &=& \frac{z_R^3}{2\, \pi^2}~
\sum_{n=1}^\infty~ (-1)^{n+1}~ \left[ 3\, K_3 (n\, z_R) + K_1 (n\, z_R)
\right]\, \sinh (n\, \pe) \pp
\eea

$~~~~~~~~~$\newpage

$~~~~~~~~~$\newpage

$~~~~~~~~~$\newpage

$~~~~~~~~~$\newpage

$~~~~~~~~~$\newpage

$~~~~~~~~~$\newpage

$~~~~~~~~~$\newpage

$~~~~~~~~~$\newpage

$~~~~~~~~~$\newpage

$~~~~~~~~~$\newpage

$~~~~~~~~~$\newpage

$~~~~~~~~~$\newpage

$~~~~~~~~~$\newpage

$~~~~~~~~~$\newpage

$~~~~~~~~~$\newpage

$~~~~~~~~~$\newpage

$~~~~~~~~~$\newpage

$~~~~~~~~~$\newpage

$~~~~~~~~~$\newpage

\newpage

\end{document}